# Responsible LLM Deployment for High-Stake Decisions by Decentralized Technologies and Human-AI Interactions

Swati Sachan
University of Liverpool
*Financial Technology*
Liverpool, United Kingdom
Swati.Sachan@liverpool.ac.uk

Theo Miller
*Chain Crunch Labs*
Kendall Square
Boston, Massachusetts, United States
Theo.Miller@chaincrunchlabs.com

Mai Phuong Nguyen
University of Liverpool
*Financial Technology*
Liverpool, United Kingdom
p.m.nguyen@liverpool.ac.uk

*Abstract*— High-stakes decision domains are increasingly exploring the potential of Large Language Models (LLMs) for complex decision-making tasks. However, LLM deployment in real-world settings presents challenges in data security, evaluation of its capabilities outside controlled environments, and accountability attribution in the event of adversarial decisions. This paper proposes a framework for responsible deployment of LLM-based decision-support systems through active human involvement. It integrates interactive collaboration between human experts and developers through multiple iterations at the pre-deployment stage to assess the uncertain samples and judge the stability of the explanation provided by post-hoc XAI techniques. Local LLM deployment within organizations and decentralized technologies, such as Blockchain and IPFS, are proposed to create immutable records of LLM activities for automated auditing to enhance security and trace back accountability. It was tested on Bert-large-uncased, Mistral, and LLaMA 2 and 3 models to assess the capability to support responsible financial decisions on business lending.

## I. Introduction

Generative Large Language Models (LLMs) have significantly advanced in computational reasoning and linguistic articulation. Numerous LLM architectures with high scalability and performance suitable for cloud and local hardware deployment have been released. However, empirical evaluation of LLMs for real-world tasks in high-stake decision domains, such as healthcare and finance, is challenging [1]. Their deployment significantly impacts large demographics, as it aims to assist domain experts in processing and interpreting vast amounts of text and images.

The technological integration of LLMs into professional settings has raised critical questions: How do we rigorously evaluate the capability of fine-tuned LLMs for real-world applications? How can we manage accountability for decisions derived from LLM? How can we address data security concerns due to the leakage of sensitive information, especially through user prompts and model poisoning attacks? A localized LLM can mitigate these issues, yet concerns over explainability and security in the local environment hinder its deployment. Due to these unresolved challenges, many firms have cautiously opted to restrict the use of Generative AI on workplace computers and avoid the deployment of local LLMs. The responsible application of LLMs for decision-making tasks requires not only a high degree of confidence in the predicted outcome but also clear interpretable explanations and robust security against adversarial threats on proprietary organizational data and model parameters. Additionally, manage the unjustifiable legal liability against human experts for incorrect outcomes who interact with these systems.

This paper presents a framework to monitor LLM performance iteratively at pre and post deployment stages by human-in-the-loop activities for continuous improvement. They fine-tune LLM with augmented data and assess the robustness of the explanation. The uncertainty and confidence in the prediction (generated text or label) are quantified to identify and redirect low-confidence instances for manual human review. For accountability attribution and security from adversarial attacks, it proposes LLM's fine-tuning with proprietary data of an organization in the local environment for specific tasks and the use of decentralized technologies such as Blockchain and Interplanetary File Systems (IPFS) to create real-time and immutable records of LLM activities, respectively. Blockchain allows permanent storage of data within a peer-to-peer computer network, whereas IPFS is an open-source, content-addressable peer-to-peer network of distributed data [2]. Only hash pointers of LLM-generated metadata files containing information such as decisions, prompts, and responses are stored in Blockchain. For scalability, IPFS (anonymized metadata) and Cloud (repositories and blobs) are used to store large and sensitive data, which is linked with Blockchain for automated auditing to track the integrity and authenticity of files over time.

## II. RELATED WORK AND REVIEW ON LLM EVALUATION

### A. LLMs Evaluations by Human and Quantifiable Metrics

The LLM performance with respect to human evaluations is commonly measured with three key metrics: percentage agreement, Cohen's kappa, Fleiss' kappa, and Scott's pi [3]. The percentage agreement measures the frequency of agreements, and Cohen's kappa evaluates the inter-rater reliability for agreements occurring by chance between LLM judges and human annotators. Fleiss' kappa extends this concept to multiple raters to assess the consistency among several human annotators and LLMs serving as judges. It generalizes Scott's pi, a metric for two annotators that presumes similar category distribution among the raters. A medical study utilized Cohen's kappa to explore GPT-4's proficiency in medical tasks [4]. A recent study quantified human alignment with models, GPT-4 and versions of Llama 2 & 3 [5]. It concluded that Scott's pi more distinctly differentiates between models compared to other metrics.

Other than manual human evaluations, perplexity is one of the most important metrics for assessing the predictive confidence of models in machine translation, speech recognition, and text generation [6]. A high perplexity score indicates a high uncertainty outcome and vice-versa. It adds interpretability in how well the model understands language and comprehends the given task by its ability to generalize across in-distribution and out-of-distribution samples. Mathematically, perplexity is exponential to cross-entropy. It is a learning objective in the training of LLMs. It measures the average log-loss for each word in the sequence. Lower cross-entropy means the model assigns higher probabilities to the actual words in the sequence. Perplexity represents the effective "number of choices" the model feels it has at each word. Entropy quantifies the inherent uncertainty or unpredictability of a distribution. In language modeling, entropy is the average uncertainty of the true distribution of words [7]. Higher entropy indicates an uncertain distribution or low confidence, and vice-versa. However, these metrics do not capture linguistic nuances such as the model's alignment with factual knowledge and real-world context. Full validation of an LLM's outputs requires interpretative analysis and domain-specific approval for a task.

### B. Explainable decisions by LLMs

The reasoning behind an outcome by LLM can be obtained by pointing out the contribution of each input feature (words and tokens). LLM outcomes can be interpreted by post-hoc explainability techniques such as LIME (Local Interpretable Model-agnostic Explanations) and SHAP (SHapley Additive exPlanations) for the contribution of each feature, such as word or token [8]. Gradient-based techniques can be used to understand the relevance of each input feature. For example, Layer-wise Relevance Propagation (LRP) redistributes relevance scores backward through the network [9]. Deep Taylor Decomposition approximates complex non-linear neural networks using the Taylor series [10]. Integrated gradients compute feature contributions by accumulating gradients by Reimann sums [11]. A recent study utilized LIME, SHAP, and integrated gradient to detect misinformation by LLMs such as Llama, Orca, Falcon, and Mistral [12]. Both LIME and SHAP are extensively used to explain BERT [13][14]. Another research utilized SHAP to create human-understandable explanations for anomaly detection in BERT, Mistral, and Llama [15].

### C. Decentralized Technology for Secure AI

The integration of decentralized technology with AI is a significant advancement towards the development of a secure and privacy-preserving environment. A recent study paired blockchain and IPFS files to record comprehensive textual explanations of XAI model decisions by generative AI tools to monitor the human usage of generated text and decisions [16]. A similar study explored auditing XAI decisions by utilizing the IPFS to address the storage constraints associated with the Ethereum network [17]. However, it did not adequately conduct robustness tests using a specific use case. In the insurance industry, blockchain was utilized for secure consent-based data sharing among various stakeholders and secure processing of medical indemnity insurance by an explainable deep learning model [18][19]. In the financial domain, a study introduced a framework to aggregate expert knowledge by fuzzy cognitive map through Blockchain platforms to create fair lending criteria [20].

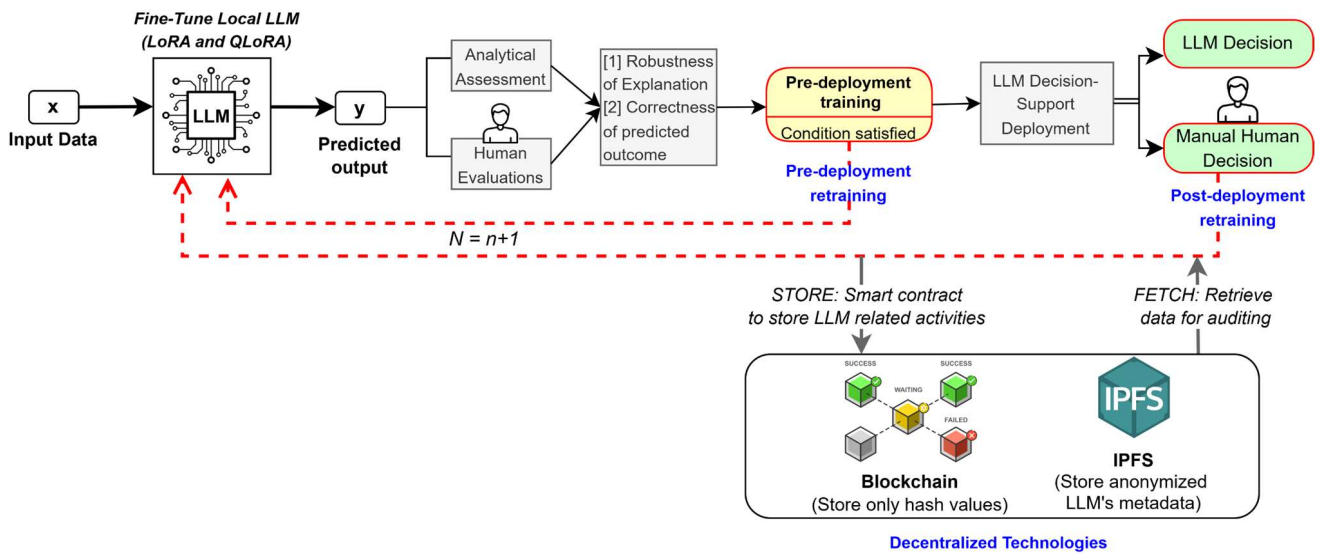

Fig. 1. Framework on Human-AI interaction and use of decentralized technologies for responsible and secure LLM deployment

## III. FRAMEWORK FOR RESPONSIBLE LLM DEPLOYMENT

This framework presents pre- and post-deployment steps to design a robust LLM-based decision-support system through active human involvement. For security and accountability management, it proposes the use of local LLMs within an organization and the utilization of decentralized technologies: blockchain and IPFS. It stores the immutable and tamper-proof records of LLM-related activities for automated and trustworthy auditing. Figure 1 illustrates the integration of human, XAI, and decentralized technology in the pre- and post-deployment stages of LLM.

### A. LLMs Evaluation by Humans and XAI

Local and open-source LLMs support data control, scalability to cloud services, and the ability to run quantized models at an acceptable speed on laptops with fewer GPUs, unlike Generative AI tools hosted by third parties. Open-source LLMs such as Llama 3, Mistral, and Bert-large-uncased are pre-trained with high-precision float32 data (32-bit float-point numbers), to store and compute the weights and activations. To reduce computational demand and memory costs on local resource-constrained devices, the quantization technique is used to convert data to lower-precision formats, such as float16, 8-bit or 4-bit integers.

Let an LLM model be denoted by $\mathcal{L}$ and its parameters by $\theta$. The input data and predicted output are denoted by $x \in X$ and $y \in Y$, respectively. An instance is denoted by $(x_i, y_i)$, such that $i \in \{1, ..., I\}$. The parameters of pre-trained LLMs $(\theta)$ are fine-tuned for specific tasks or decisions $(y)$ with domain-specific data $(x)$, such as finance and medical documents, to improve accuracy. To train LLM, the fine-tuning objective is to optimize the parameters by a loss function (usually cross-entropy) to minimize the discrepancy between the actual output $(y)$ and predicted output $(\hat{y}(x, \theta))$ as shown below:

$$\theta^* = arg\ min_\theta\ \mathcal{L}_{loss}(y, \hat{y}(x, \theta)) \qquad (1)$$

Full fine-tuning of LLMs requires a substantial amount of time and high-performance GPUs, depending on data size (number of samples and token length) and model size (number of parameters). To address this issue, Parameter-efficient Fine-tuning (PEFT) techniques such as Low-Rank Adaptation of Large Language Models (LoRA) and Quantized Low-Rank Adapters (QLoRA) are developed to optimize and enhance inference efficiency on local computers. LoRA freezes the pre-trained model weights and introduces trainable rank decomposition into layers of the transformer architecture [21]. QLoRA further reduces computational demands by quantifying the precision of the weight parameters to 4-bit precision [22].

Fine-tuning LLM aims to store domain-specific factual knowledge in their parameters. At the pre-deployment stage, humans conduct fact-checking to evaluate the quality of the outcome and the robustness of the explanation of a decision using XAI techniques. The samples evaluated by humans are used as augmented data to refine the LLM iteratively. Following are pre- and post-deployment steps

- STEP 1: Sample Selection for Human Evaluation

A model's accuracy is evaluated by the F1-score and Matthew's correlation coefficient (MCC) to assess its precision and recall. However, uncertainty and confidence in the prediction of language models are measured by perplexity and entropy, denoted by $ppl$ and $H$, respectively.

The perplexity of fixed-length models should be evaluated with a sliding-window strategy [23]. Instead of estimating the perplexity of each word in a sequence, the sequence is broken into disjoint chunks of words, and then the decomposed log-likelihoods of each segment are added independently. Entropy measures the general uncertainty and information diversity within the data analyzed by the LLM. The entropy and perplexity of each instance in the test set are estimated to find a threshold to select a set of samples for human evaluation to be used as augmented data for reiterative fine-tuning. The entropy value ranges from [0,1]; however, the perplexity value is not standardized. It depends on text data size and context length. The entropy and perplexity of a single instance and entire dataset (training or test set) are:

$$H = \begin{cases} h_i = -p(x_i)\log p(x_i), \text{Single Instance} \\ H_X = -\sum_{i=1}^{I} p(x_i)\log p(x_i), \text{ Entire dataset} \end{cases} \qquad (2)$$

$$ppl = \begin{cases} ppl_i = e^{-p(x_i)\log q(x_i)}, \text{Single Instance} \\ ppl_X = e^{-\sum_{i=1}^{I} p(x_i)\log q(x_i)}, \text{ Entire dataset} \end{cases} \qquad (3)$$

The exploitation and exploration strategy are applied to each instance of the test set, which is sorted in increasing order of entropy and perplexity. A set of samples was selected for human evaluation is denoted by $S$. It contains samples from three entropy and perplexity regions: $s_{HC}$ a set of high-confidence $(HC)$, $s_{MC}$ a set of moderately uncertain $(MC)$, and $s_{LC}$ a set of low-confidence $(LC)$, or highly uncertain samples with the perplexity below a threshold value (example: below 25th percentile), as shown by the below expression:

$$S = \begin{cases} s_{HC}: [0 < h_i \leq 0.5\ or\ ppl_i \geq ppl^{THR}] \\ s_{MC}: [0.25 < h_i < 0.75\ and\ ppl_i \geq ppl^{THR}] \\ s_{LC}: [0.75 \leq h_i \leq 1\ and\ ppl_i \geq ppl^{THR}] \\ \text{here, } s_{HC} \cap s_{MC} \cap s_{LC} = \emptyset;\ s_{HC}, s_{MC}, s_{LC} \in S \end{cases} \qquad (4)$$

The cardinality (the total number of instances) of the selected sample set is denoted $|S|$. There is an overlap in the region between the moderately uncertain group and the other two groups. However, sample duplication is avoided, $s_{HC} \cap s_{MC} \cap s_{LC} = \emptyset$. A smaller proportion of samples are selected from the high-confidence region compared to the moderately uncertain and low-confidence regions, as expressed below:

$$|S| = (HC_\%\ |s_{HC}|) + (MC_\%\ |s_{MC}|) + (LC_\%\ |s_{LC}|)\quad (5.1)$$

$$HC_\% \ll MC_\% \ll LC_\% \qquad (5.2)$$

- STEP 2: Human Evaluation Score

Fleiss' kappa is utilized to measure the reliability of agreement between a fixed number of human evaluators and multiple categories [3,4]. Evaluators judge the correctness of the model's predictions (accuracy) in two categories: $\{agree, disagree\}$. They also evaluate the quality of reasoning provided by XAI methods in four categories: $\{poor, moderate, good, excellent\}$. For each sample $(s \in S)$, the Fleiss' kappa score is obtained by human evaluators to assess decision accuracy $(y)$ and the clarity of the explanation by XAI technique $(g)$, denote by $\kappa_y^s \in [0,1]$ and $\kappa_g^s \in [0,1]$, respectively. The value $\kappa = 1$ indicates perfect agreement among the evaluators and vice-versa. It is derived from the observed agreement $(A_o)$ and estimated agreement $(A_e)$ between evaluators:

$$\kappa_y^s = \frac{A_o^{s,y} - A_e}{1 - A_e}; \quad \kappa_g^s = \frac{A_o^{s,g} - A_e}{1 - A_e} \quad (6)$$

- STEP 3: Lexical Robustness of Explanation

Recognizing the robustness of explanations generated by post-hoc XAI techniques such as LIME, SHAP, and gradient-based methods is crucial at the pre-deployment stage. Developers use this phase to monitor the AI system's behavior and performance to ensure that the LLM functions as expected in controlled environments before it is released into real-world applications. Simultaneously, domain experts evaluate the explanations to verify the logic and reasoning behind decisions to ensure their alignment with domain-specific facts and knowledge.

At the post-deployment stage, the focus shifts towards operational use, where human experts utilize the explanations to inform their final decisions. A clear and understandable explanation helps to attribute responsibility and accountability in the event of an adversarial decision to protect human experts from unjust liability claims.

The stability of explanation generated by XAI techniques, denoted as $g \in \{1, \ldots, G\}$, is examined by a change in the saliency score for a word $w$ in a given instance $x_i$ and its local perturbed version $x_i'$. The perturbed version is generated without losing the actual meaning and context of the original instance $x_i$. Additionally, some lexical similarities between $x_i$ and $x_i'$ is incorporated to test the resilience of the XAI's outputs. The stability metric, denoted by $\mathrm{E}_{i,g}(w)$, is:

$$\mathrm{E}_{i,g} = \frac{\sum_{w=1}^{W} \varepsilon_{i,g}(w)}{W}; \; w \in x_i \quad (7.1)$$

$$\varepsilon_{i,g}(w) = \sum_{g \in G} \left\| x_{i,g}(w) - \left( x_{i,g}'(w) \; \phi_{i,g}(w) \right) \right\| \quad (7.2)$$

Here, $x_{i,g}(w)$ and $x_{i,g}'(w)$ is the saliency score for a word $w$ in instance $x_i$ and $x_i'$ obtained from an XAI technique $g$, respectively. The presence of the word $w$ or its contextually related terms (e.g., synonyms) in the perturbed instance is pointed by indicator function $\phi_{i,g}$. A complete evaluation of an XAI technique by analytical method (stability metric) and human evaluation (Fleiss' kappa score) for a selected set of samples is:

$$E_g = \beta_1 \left( \frac{\sum_{\forall i \in S} \mathrm{E}_{i,g}}{|S|} \right) + \beta_2 (k_g) \quad (8)$$

Here, $\beta_1$ and $\beta_2$ are weighted factors to combine the human and analytical evaluation. For equal contribution, $\beta_1 = \beta_2 = 0.5$ such that $E_g \in [0,1]$.

- STEP 4: Iterative Retraining in Pre-Deployment Stage

The LLM model undergoes $N$ times iterative retraining during the pre-deployment phase. The retraining continues until the model's performance on a set $S$ of samples meet specific thresholds. The following is the retraining termination condition based on the higher human agreements and low-uncertainty thresholds:

$$(k_y \geq k_y^{THR}) \land (E_g \geq E_g^{THR}) \land (H \leq H^{THR}) \land (ppl \leq ppl^{THR}) \quad (9)$$

$$N = \begin{cases} n+1, \text{if condition (9) is TRUE} \\ 0, \text{otherwise} \end{cases} \quad (10)$$

In Conditional Expression (9), ∧ represents AND logic.

- STEP 5: Post-Deployment Stage

At the post-deployment stage, the LLM is launched for real-world tasks after undergoing iterative testing at the pre-deployment stage. However, model improvement does not stop at the post-deployment stage. During operation, instances processed by the LLM with certainty above a predefined threshold from the last iteration are accepted. However, cases with high uncertainty are forwarded to human experts for manual review. The expert reviewed samples can be used to fine-tune the model either periodically or after significant drops in overall decision accuracy. The sample selection for human data augmentation is conducted by the technique presented in Step 1.

B. *Accountability and Security by Blockchain and IPFS*

Successful integration of AI with decentralized technologies requires adaptations to meet a firm's stakeholder demands for compliance with data protection laws such as the General Data Protection Regulation (GDPR), which strictly restricts the permanent storage and exposure of personal information.

A hybrid on-chain and off-chain data storage strategy is employed for scalability and cost-efficiency, shown in Table I. Off-chain data is stored in the Cloud and IPFS. LLM's metadata, parameters, and internal organizational data are stored in the Cloud; however, it is frequently accessed by internal stakeholders such as developers and domain experts. It is vulnerable to unintentional tampering and malicious attacks. Therefore, IPFS is paired with Cloud and Blockchain as a scalable solution for storing large off-chain data and recovery of original untampered files if Cloud storage is compromised or accidentally altered.

A smart contract (or chain code) records the cryptographic hash value of a metadata file, a unique hexadecimal string representing a digital fingerprint of a data file. SHA-256 (256-bit or 32-byte string) is a popular hashing function. A metadata file contains information about inputs (prompts) and outputs (responses) from LLM, human expert ID, model name, expert's final decision, LLM's decision explanation by a robust XAI technique, static IP address of the workplace computers, shown in Figure 2.

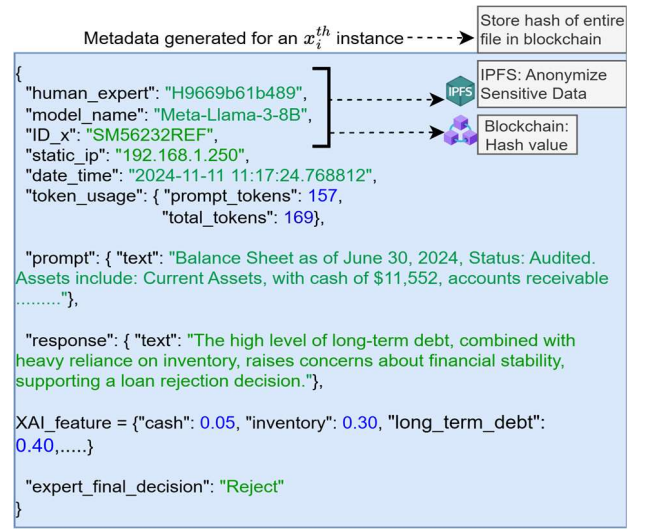

Fig. 2. An example of metadata file for on-chain and off-chain

TABLE I. ON-CHAIN AND OFF-CHAIN DATA STORAGE

| On-Chain: $\beta$ (Blockchain Network) | *Separate Hash*: Human expert's ID, LLM model ID the static IP address of the authorized workplace computer |
| --- | --- |
| | *Combine Hash*: A hash of a metadata file of an instance containing the prompt and response stored in cloud |
| Off-chain: $\alpha$ | *Cloud*: Customer raw data (documents) and metadata related to the LLM-based decision-support system |
| | *IPFS*: Anonymized metadata file |
| | *Key Storage*: Blockchain node identifiers, human expert and other administrative IDs |

A human-readable metadata file in JSON format, denoted as $\phi_x$, is stored in the cloud and its anonymized version for storage on the IPFS is denoted as $\phi'_x$. The hash values of the files stored in the cloud and on IPFS are represented by $h_x$ and $h'_x$, respectively. These hash values are combined into a single JSON object and encoded in BASE64 format for 256-bit slots compatible structure in blockchain and IPFS:

$$D_\beta^x \leftarrow BASE64(JSON\{h_x, h'_x\}) \qquad (11)$$

Any alteration of the metadata and questionable decisions indicate deviations from established ethical standards and allow tracing accountability back to humans (domain experts, developers, or external malicious actors) for inconsistent adversarial decisions. The tampering state ($\tau$) indicates tampering, if the re-calibrated hash value of files stored in the cloud and IPFS at an auditing time $t$ do not match the original hashes stored on the blockchain. The tampering condition is:

$$\tau_t = \begin{cases} 0, & h_x \neq Cloud\_h_\alpha^x \text{ and } h'_x \neq IPFS\_h_\alpha^x \\ 1, & \text{other wise} \end{cases} \qquad (12)$$

## IV. TEST RESULTS

### A. LLMs for Financial Underwriting Decisions

The proposed framework is applied to develop a decision-support system based on LLMs to assist human underwriters in making funding decisions for small business loans. Five LLM model architectures: Bert-large-uncased 340M, Mistral 7B, LLama2 7B, LLama2 13B, and Llama3 8B, were tested with a corpus of 8.4k financial statement reports. Each report contains an average of 200 tokens. The data was divided into four parts: 4.8k samples for training and three sets of 1.2k samples for validation and iterative improvement at the pre-deployment stage ($N = 3$). The experiment was conducted on Apple M3 with a 16-core GPU. To optimize memory usage, the Llama and Mistral LLMs were quantized by QLoRA, whereas LoRA quantized Bert-uncase to host LLMs in the local computer. All models were fine-tuned with lower precision floating-point for five epochs. The memory requirement for inference, fine-tuning, and performance metrics are detailed in Table II. Among all models, Llama3 8B demonstrated the best performance, and Bert-large-uncase had the worst performance after three iterative improvements in collaboration with human experts, as detailed in Table II. LLama2 13B exhibited the next best performance, followed by a similar performance of Llama 2 7B and Mistral 7B.

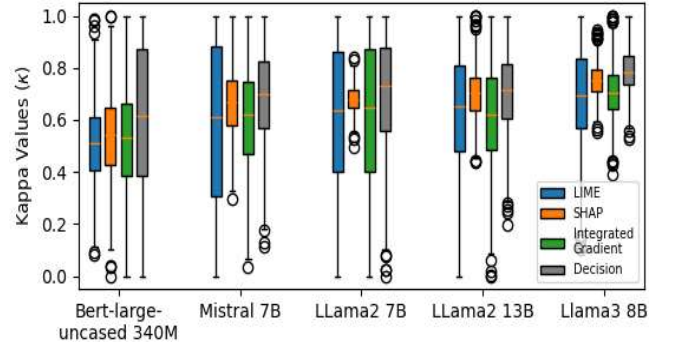

Fig. 3. Importance of features in financial statements by SHAP. Red indicates the word importance for the "reject" decision, and green for the "fund" decision.

The decision-making capabilities of the models and their explanatory reasoning for a given instance by LIME, SHAP, and Integrated Gradient were assessed by three human experts across 70 samples over three iterative cycles. Experts evaluated the quality of explanations by examining the heatmap of feature importance for each input decision, as shown in Figure 3. The Fleiss' kappa score ($\kappa \in [0,1]$) reflects the stability and human understanding of explanation by XAI. The Kappa score was highest for SHAP, with the lowest variance in disagreement compared to LIME and integrated gradient, as shown in Figure 4. Similarly, the agreement on the quality of decisions made by Llama 3 was the highest, with the lowest variance among all tested models.

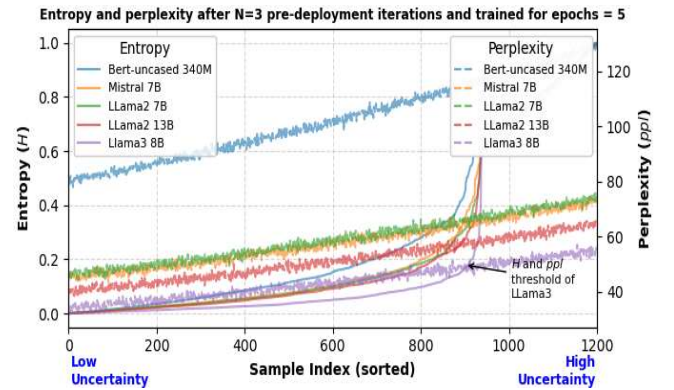

Fig. 4. Fleiss' kappa score by human assessment of decisions given by LLM and the robustness of explanations provided by three XAI techniques

Figure 5 illustrates the 1200 samples sorted in increasing order of perplexity and entropy, following three pre-deployment iterations of improvement in LLMs. If Llama 3 is chosen due to its strong performance and high level of human agreement, then outcomes with entropy below 0.164 and perplexity under 47.824 can be accepted when processed by the LLM. Cases that exceed these thresholds should be redirected to domain experts for further review.

Fig. 5. Sample sorted in increasing order of entropy and perplexity, that is from low uncertainty (high confidence) to high uncertainty (low confidence)

TABLE II.  LLM MODEL REQUIREMENTS AND EVALUATION METRICS AFTER N=3 ITERATIONS BETWEEN HUMAN AND LLM

| Model | Memory (Inference) | Memory (Fine Tuning) | Precision | Recall | F1 Score | MCC | Mean Perplexity | Mean Entropy |
|---|---|---|---|---|---|---|---|---|
| Bert-large-uncased 340M | 12 GB | 3 GB (16bits) | 0.180 | 0.151 | 0.164 | -0.543 | 80 | 3.25 |
| Mistral 7B | 14 GB | 4 GB (4bits) | 0.711 | 0.691 | 0.700 | 0.410 | 45 | 1.8 |
| LLama2 7B | 28 GB | 6 GB (4bits) | 0.715 | 0.601 | 0.653 | 0.366 | 46 | 1.9 |
| LLama2 13B | 52 GB | 12 GB (4bits) | 0.792 | 0.801 | 0.795 | 0.588 | 40 | 1.4 |
| Llama3 8B | 16 GB | 6 GB (16bits) | 0.812 | 0.821 | 0.816 | 0.630 | 34 | 1.1 |

## B. AI Accountability by Decentralized Technologies

Two types of LLM activities are recorded on the blockchain: loan application decisions and updates to LLM parameters over time. Auditing these activities for accountability and detecting tampering or malicious attacks was conducted on two blockchain networks: public (Polygon zkEVM) and private (Hyperledger Fabric v2). The demand to update information on a blockchain platform escalates with an increase in the number of users and nodes (servers). Hyperledger Fabric exhibited 7.56% higher throughput and 6.15% lower latency compared to Polygon; however, this performance difference is marginal compared to the Ethereum Layer 1 solution. In this experiment, the network traffic (number of transactions) was uniformly distributed. Transactions were sent through JSON-RPC with a maximum batch size of ten and a limit of 10 MB (Megabytes) per batch in both networks.

The efficiency of the audit process was assessed by introducing random alterations in 2% to 18% of the files stored in off-chain mediums. The goal was to ensure that the recomputed hash of off-chain storage mediums (IPFS and Cloud) matched the corresponding hashes permanently stored on the blockchain to pass the automated audit successfully. The completion time of the audit process grows with the increase in the number of tampered files in Polygon and Hyperledger Fabric. Figure 6 illustrates the audit time analysis of Llama 3 8B.

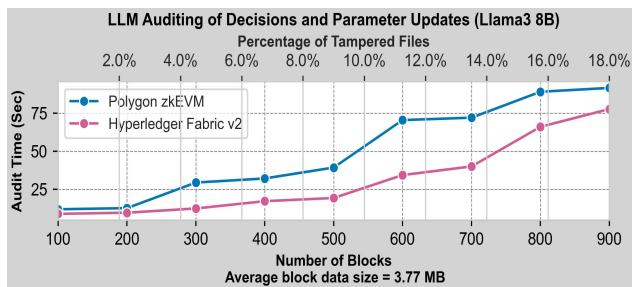

Fig. 6. Audit Trails Analysis of Llama 3 8B LLM model in Polygon and Hyperledger Fabric

## V. CONCLUSION

This research presents the significant potential for effective governance and responsible deployment of LLMs in real-world settings by integration of decentralized, cryptographically secured technologies and interactive human-AI activities. The proposed framework is applied to assist human underwriters in making informed, high-stakes financial decisions for small business lending through detailed analysis of financial reports. The human underwriters and developers worked collaboratively at the pre-deployment stage to improve the LLM performance with stable explanations. Automated auditing of LLM activities recorded on decentralized platforms promotes the responsible use of AI algorithms and reduces inconsistencies in tracking accountability in the event of adversarial decisions.